\documentclass[twocolumn,aps,prl,superscriptaddress,showpacs,hyperref]{revtex4}
\usepackage{graphicx}
\usepackage{mathptm}

\begin{document}

\title{Stabilization of the output power of intracavity frequency-doubled lasers}

\author{F.~Lange}
\affiliation{Department of Energy and Semiconductor Research,
Faculty of Physics, University of Oldenburg, D-26111 Oldenburg}

\author{T.~Letz}
\affiliation{Department of Energy and Semiconductor Research,
Faculty of Physics, University of Oldenburg, D-26111 Oldenburg}
\affiliation{Max-Planck Institute for Complex Systems, Dresden,
Germany}

\author{K.~Pyragas}
\affiliation{Department of Energy and Semiconductor Research,
Faculty of Physics, University of Oldenburg, D-26111 Oldenburg}
\affiliation{Semiconductor Physics Institute, LT-2600 Vilnius,
Lithuania}
\author{A.~Kittel}
\email{kittel@uni-oldenburg.de}
\affiliation{Department of Energy
and Semiconductor Research, Faculty of Physics, University of
Oldenburg, D-26111 Oldenburg}
\date{28 August 2003}

\begin{abstract}
Intracavity frequency-doubled solid-state lasers exhibit intensity fluctuations of their light
output, which are cause by nonlinear dynamical processes. Up to now, there are different solutions
to this problem, but they reduce the output power, increase the size of the laser and/or make them
more complicated to assemble. One focus of current research in nonlinear dynamics is derivation of
control strategies from mathematical models and their experimental realization. We suggest a method
to stabilize the output power by means of an electronic feedback of the emitted infrared light
intensity to the pump power. First we show the theoretical predictions of a recently published
stability analysis of a rate equation model with feedback. The presented experimental observation
show systematic deviations from theory. This makes it necessary to refine the model to explain the
deviations. The refinement has direct impact on the improvement of the feedback loop and,
therefore, on the application of such a control scheme.
\end{abstract}

\maketitle

Intracavity frequency-doubled solid-state lasers as light sources in the visible spectral range
have a high technical potential because of their efficiency as well as their favorable ratio of
compactness and output power \cite{text}. Indeed, the intracavity frequency conversion design makes
the nonlinear processes highly efficient, however, under multimode operation disadvantageous
dynamical instabilities of the system, commonly known as the \emph{green problem}, are often
observed in such kind of lasers. The instabilities originate from a coupling of the individual
resonator modes due to the processes of sum frequency generation inside the nonlinear crystal and
spatial hole burning within the laser-active media. Any kind of fluctuation of the output intensity
can be a serious obstacle to many types of technical applications. Therefore, many people in
industry and science are working on the stabilization of such kind of lasers.

Since the first scientific discussion of the green problem by Baer \cite{baer1986} a number of
proposals were made concerning the suppression of the fluctuations by means of different optical
modifications of the resonator (e.g. ring resonator, mode selecting etalon)
\cite{anthon1992,james1990a,oka1988}. An alternative approach was proposed by Roy et al.
\cite{roy1992}. Leaving the optical setup of the system unchanged they exploited the existence of
unstable steady states to perform a stabilization of the laser output intensity on a periodic orbit
utilizing the so called occasional proportional feedback, known as OPF. As far as the application
is concerned, it is more desirable to achieve a constant output intensity instead of a stable
oscillating one. Nevertheless, it was demonstrated that it is possible to control the output
intensity of the laser by a pure electronic feedback in principle.

In this letter theoretical aspects and experimental results of an alternative control method are
presented. From a rate equation model \cite{james1990a} we derived a control scheme
\cite{pyragas1999} that is able to stabilize the laser output by feeding back a control signal
$u(t)$. The signal $u(t)$ can be derived from the sum intensities of the two orthogonal directions
of the linear-polarized infrared light intensities which are emitted by the laser. More precisely,
$u(t)$ has the form
\begin{equation}
    u(t)=k_x (s_x^0 - s_x(t)) + k_y (s_y^0 - s_y(t)),  \label{cs}
\end{equation}
where the light intensities $s_x$ and $s_y$ correspond to the sum of the individual modes polarized
in the orthogonal directions $x$ and $y$, respectively. The directions $x$ and $y$ correspond to
the polarization eigenvectors of the laser cavity and depend on the birefringence and the length of
the crystals inside the laser cavity. $s_x^0$ and $s_y^0$ are the time-independent sum intensities
at the steady state and $k_x$ and $k_y$ are two control parameters. First experimental results on
such kind of stabilization have been reported in \cite{alex2001}.

The approach to form the feedback signal like Eq.(\ref{cs}) results from an analysis of rate
equations, which consists of two ODEs for each individual lasing mode. Each set of ODEs for the
individual laser modes incorporates a set parameter which are in general different from each other.
The parameters are experimentally not accessible, therefore, for simplicity (w.l.o.g.) and to treat
the system analytically we have chosen for all the laser modes the same set parameters. It can be
shown, that the equations modelling the dynamics of the sum intensities $s_x$ and $s_y$ break off
from the total system of equations describing the dynamics of each individual mode intensities in
vicinity of the fixed point \cite{pyragas1999}. The stability of each individual laser mode
intensity is reduced to the problem of the stability of the sum intensities $s_x$ and $s_y$. By
using a control signal $u(t)$ in accordance with Eq.(\ref{cs}) and choosing proper values for the
control parameters $k_x$ and $k_y$ one can theoretically achieve and maintain the stability of the
fixed point intensities as shown by Pyragas et al. \cite{pyragas1999}. In the following we will
give evidence that this control strategy also works experimentally in agreement with the
theoretical predictions.

\begin{figure}
\centerline{\includegraphics[width=7cm,angle=0]{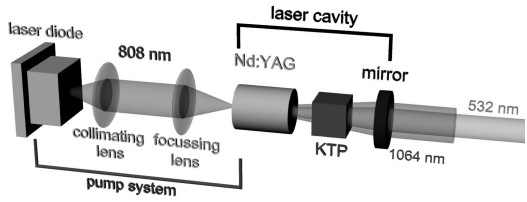}}
    \caption{Schematic drawing of the intracavity frequency doubled Nd:YAG laser pumped by a
    laser diode.}
    \label{laser}
\end{figure}

The experimental setup of our intracavity frequency doubled Nd:YAG laser is shown schematically in
Fig.~\ref{laser}.
A collimated beam of an infrared pump laser diode is focused on one end of a cylindrical Nd:YAG
crystal. The optical resonator is formed by the planar HR coated surface of the Nd:YAG-crystal and
a concave output mirror. A type II phase-matched KTP crystal is placed inside the cavity and serves
as the optical nonlinear element for generation of radiation at the second harmonics at
$\lambda=$532nm. Due to the optical density of the Nd:YAG and the KTP crystal of about 1.82 and
1.8, respectively, the optical length of the resonator is about 43mm.

The infrared and green spectral components of the total output intensity $S$ was separated by a
spectral selective beam splitter and the sum intensities $s_x$ and $s_y$ at 1064nm were separated
by polarizing beamsplitter. Special care was taken to operate the laser in the TEM$_{00}$ mode of
the laser resonator. In general, with increasing pump power the number of oscillating longitudinal
resonator modes increases. The dependence of the intensities $I_k$ of the individual modes on the
pump power $P_{LD}$ is complicated and is mainly governed by the process of spatial hole burning
\cite{tang1963}. Spatial hole burning originates from competition of modes for the same locally
exited states inside the active media and results for instance in the fading of a laser mode with
increasing pump power in favor of an other one. In order to decide which kind of nonlinear mode
coupling takes place inside the KTP crystal, i.e. sum frequency generation or frequency doubling,
the mode spectrum at $\lambda=$532nm can serve as a good indicator. The ratio between sum frequency
generation and frequency doubling can be controlled by rotating the main axes of the birefringent
KTP crystal \cite{james1990a}. For the following experiments a constellation with only sum
frequency generation was chosen.

\begin{figure}
\centerline{\includegraphics[width=7.5cm,angle=0]{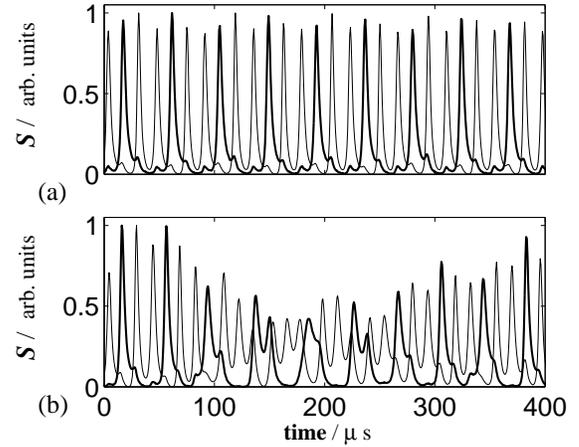}}
    \caption{Measured time series of the infrared sum intensities $s_x$, bold line, and $s_y$, light line,
    polarized in $x$- and $y$-direction, respectively, at two different relative pump powers $w$
     ((a) $w=1.89$ and (b) $w=2.07$).}
\label{timeseries}
\end{figure}

Two photodetectors were used in order to detect the temporal changes of the sum intensities of the
modes polarized in the two directions. The voltage signals of the detectors saved in a computer by
means of an analog-to-digital converter (100MSa, 12bit). Two typical time traces are shown in the
diagrams (a) and (b) of Fig.~\ref{timeseries}.
The fluctuations displayed in Fig.~\ref{timeseries}(a) are regular and were observed at a relative
pump power of $w=1.89$. The mode configuration was $\it [1,4]$ (one mode polarized in $x$ direction
and four modes polarized in $y$ direction). For the measurement in Fig.~\ref{timeseries}(b) the
relative pump power $w$ was increased to a value of 2.07. All other system parameters remained
unchanged. The corresponding mode configuration has changed to $\it [1,5]$. Irregular fluctuations
can be seen. Note that the time traces of the sum intensity $s_x$ correspond to the dynamical
behavior of a single mode in both diagrams and that the oscillations of $s_x$ and $s_y$ are out of
phase indicating a mode competition process, known as the antiphase dynamics. For increasing pump
power, the characteristic frequencies of the dynamics increase, following a square root behavior.
The characteristic frequency is below 200kHz for all the measurements presented here.

\begin{figure}
\centerline{\includegraphics[width=5cm,angle=0]{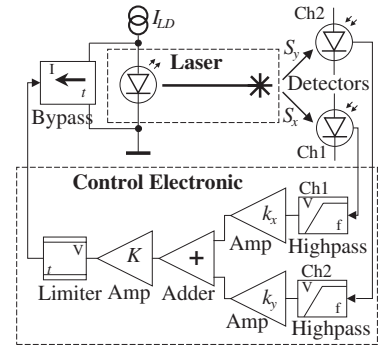}}
    \vspace{0cm}
    \caption{Schematic diagram of the control feedback loop, to perform the stabilization
    by proportional feedback}
\label{electronic}
\end{figure}

The electric circuit used to implement the proportional feedback control is displayed schematically
in Fig.~\ref{electronic}.
Instead of calculating the actual difference $s_*-s_*^0$ (where $*$ has to be $x$ or $y$) as
written in Eq.~\ref{cs}, which is complicated, because in the experiment the values of the fixed
point intensities $s_*^0$ are not known, we use a trick to estimate the difference by using high
pass filtered signal. This approximation is valid because the fixed point intensity is equal to the
time average of the intensity fluctuations. This procedure is discussed in more detail in
\cite{pyragas1999}. For this purpose we have used two high-pass filters with a cutoff frequency of
about 1kHz. Subsequently, the two signals are weighted with a variable gain amplifier ($-1\le
K_{x,y}\le1$) and afterwards added up. The gain factors $g_x$ and $g_y$ of the amplifiers were
adjusted by means of digital potentiometers and controlled remotely via a personal computer. A
third amplifier with variable gain $K$ was used. Overall we end up with a voltage signal $V_c$,
given by $V_c = k_x(V_x-V^{dc}_x) + k_y(V_y-V^{dc}_y)$, where $V_*-V_*^{dc}$ is approximated by the
high pass filtered signal from the detectors. Additionally, a limiter was used to restrict the
distortions which were applied to the laser diode at high pump powers. This is only important
during the time, where the control is switched on and the dynamics hasn't reached the vicinity of
the fixed point. In order to feed the control voltage back into the laser, the current of the laser
diode was modulated by a bypass connected in parallel to the laser diode. The -3dB bandwidth of the
control circuit was about 300kHz and the phase shift was lower than 30 degrees at the highest
characteristic frequency of the laser.

\begin{figure}
\centerline{ \includegraphics[width=4cm,angle=0]{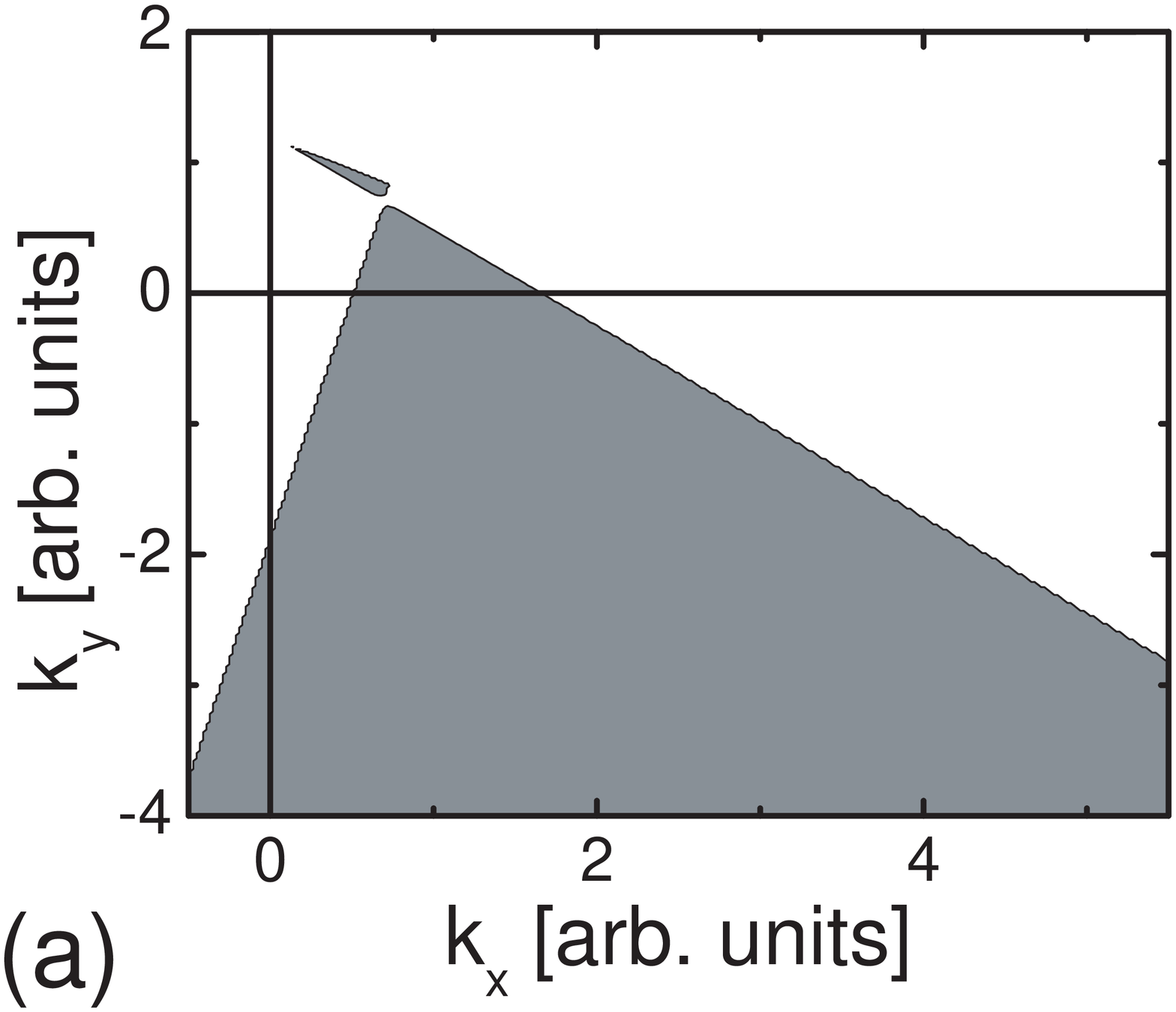} \hfill
    \includegraphics[width=4cm,angle=0]{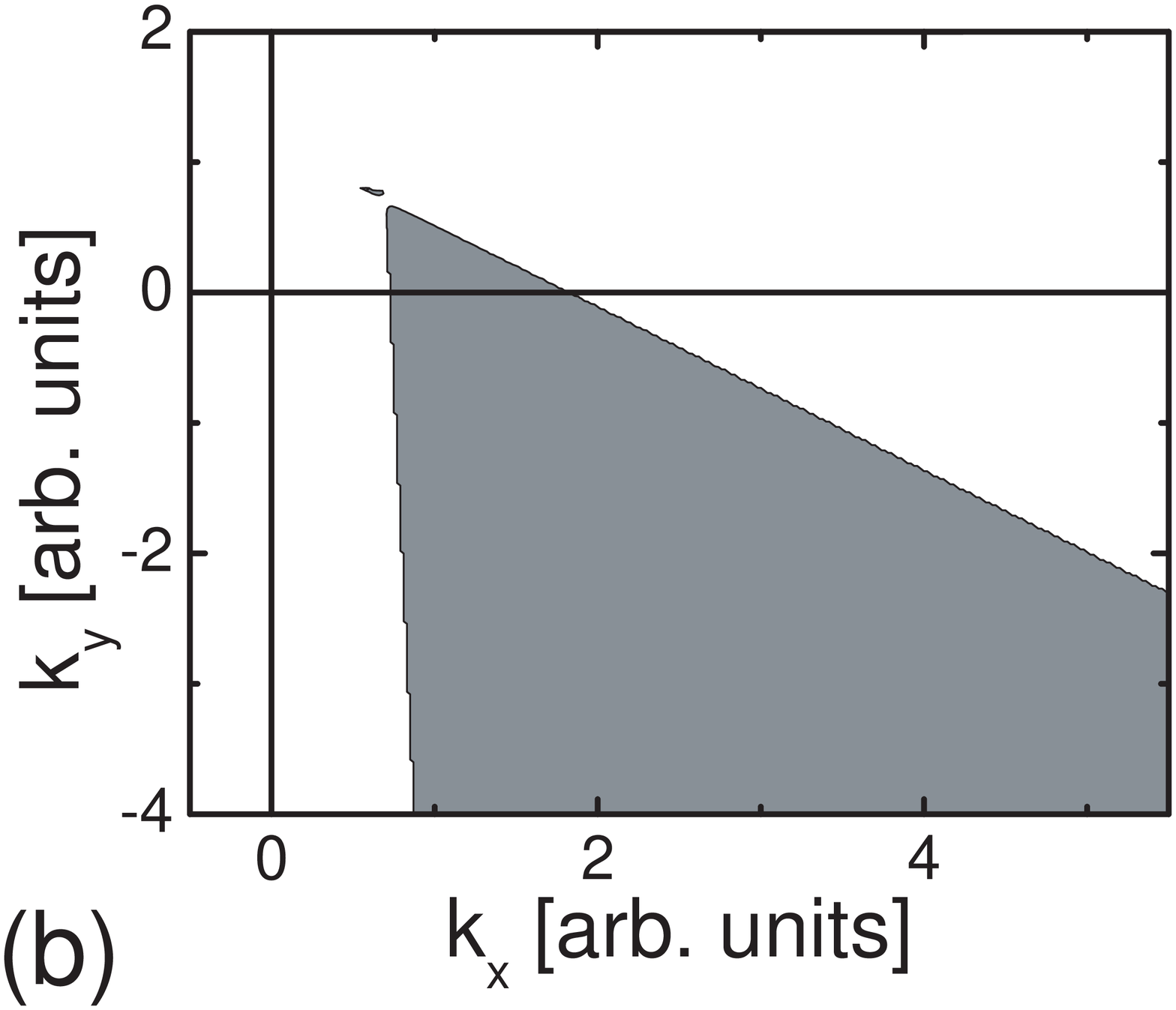}}
    \caption{Theoretical domains of stability (dark region) in the $(k_x,k_y)$ plane obtained for
    proportional feedback control. (a) At a fixed mode configuration of $\it [1,2]$ and a relative pump
    power of $w=1.38$. (b) The mode configuration was $\it [1,4]$ and the relative pump power was $w=2.07$.}
    \label{stabdomtheo}
\end{figure}

In Fig.~\ref{stabdomtheo} the theoretical domains of stability are displayed as shaded areas in the
plane of the control parameters $(k_x,k_y)$ according to theory \cite{pyragas1999}. The data were
taken for two different exemplary situations with values of the relative pump power of $w=1.38$ and
$w=2.07$ and mode configurations $\it [1,2]$ and $\it [1,4]$ in case of Fig.~\ref{stabdomtheo}(a)
and Fig.~\ref{stabdomtheo}(b), respectively. The parameters have been chosen to be as close as
possible to the performed experiments. Please note, that the mode configuration is the only
possibility in the model to have a handle on the mean sum intensity ration of the two polarization
directions, because in the model all parameters for the different modes are chosen to be the same
for simplicity. In the experiment it is more likely that the asymmetry of the intensities is caused
by different sets of parameters for the different modes but they are experimentally not accessible.
Most remarkable is the wedge-like shape of the two regions of stability of infinite size for larger
$k_x$ and more negative $k_y$ values. Increasing the pump power, the angle between the two border
lines of the region decreases.

\begin{figure}
\centerline{\includegraphics[width=8.0cm,angle=0]{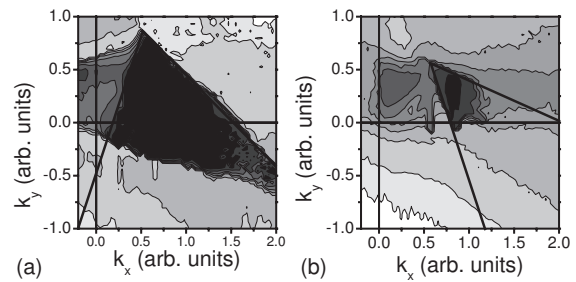}}
    \caption{Experimental domains of stability in the $(k_x,k_y)$ parameter plane obtained for
    proportional feedback control (a) at a relative pump power $w=1.38$ and the mode configuration $\it [1,1]$
    as well as (b) at a relative pump power $w=2.07$ and the mode configuration $\it [1,4]$. The
    color codes the standard deviation of the laser intensity output (the darker the color, the lower
    the standard deviation, i.e., black is closed to zero deviation).
    The two straight bold lines in both of the diagrams are added to guide the eye.}
\label{stabdomexp}
\end{figure}

In Fig.~\ref{stabdomexp} two measured domains of stability observed at different relative pump
powers are displayed.
We varied the two parameters $(k_x,k_y)$ systematically by taking data at each combination of a
hundred values in each direction and, therefore, scanned the parameter plane. As a measure for the
stability, the standard deviation of the sum intensity $s_y$ was chosen, which directly relates to
the amplitude of the fluctuations. Darker regions correspond to lower values of the standard
deviation. Thus, the black regions mark the domain with nearly no fluctuations, that is, the domain
in which the laser has a constant output power. For the two measurements shown in
Fig.~\ref{stabdomexp} only the pump power was changed, all other system parameters were kept
constant. For the measurement corresponding to Fig.~\ref{stabdomexp}(a) the relative pump power $w$
was 1.38. The laser oscillated in two orthogonal modes, $\it [1,1]$ and the output intensity showed
weak periodic fluctuations. For the second measurement corresponding to Fig.~\ref{stabdomexp}(b),
the pump power was increased to a value of $w=2.07$. At this pump power the fluctuations were
highly irregular with a dynamic range of 100\%. This means, the laser switched on and off. The mode
configuration was $\it [1,4]$. As one can see in Fig.~\ref{stabdomexp} the output power could be
stabilized at different pump powers and different mode configurations. The accordance of the
wedge-like shape of the experimental and the theoretical domains of stability is interesting. Also
the tendency of the angle of the wedge (indicated in Fig.~\ref{stabdomexp} by two black lines) to
decrease with increasing pump power is nicely reproduced. But in contrast to the theoretical
predictions, the domains of stability are finite and shrink in general due to increasing pump power
to zero (in the example presented here, the stabilization fails entirely for $w>2.2$). An extension
of the fundamental model didn't give the right explanation \cite{pyragas2000}.

To understand this behavior we have to consider an important issue, which we have neglected up to
now. The region of stability was calculated using linear stability analysis by only looking at the
real part of the eigenvalue without taking into consideration the imaginary part of the eigenvalue.
Looking at the imaginary part of the eigenvalue we notice that the imaginary part increases if
$k_y$ becomes more and more negative. From the experimental point of view it is obvious that the
feedback signal is not modelled correctly. In any experimental situation the bandwidth of the
feedback loop is restricted which was not considered in the model --- in the model the bandwidth is
unlimited. This fact influences not only the region of stability in the $(k_x,k_y)$ parameter
plane, but also the behavior of the stability region under variation of the pump power. As we
already mentioned above, the characteristic frequency of the fluctuations of the laser output
intensity grows by increasing the pump power and, therefore, the region of stability also shrinks
for a certain bandwidth of the feedback loop if the pump power is increased.

For the stability of the system it is most important to take an upper cutoff frequency into
account. The model has to be modified by an additional degree of freedom introduced by the low pass
characteristics of the feedback loop $u_{l}$. This can be modelled by adding an additional
differential equation to the rate equations introduced in \cite{pyragas1999}. The additional
equation reads as follows:
\begin{equation}\label{low_pass}
    \frac{du_{l}}{dt}=\omega_c(k_x (s_x^0 - s_x(t)) + k_y (s_y^0 - s_y(t))-u_{l}),
\end{equation}
\begin{figure}
\centerline{ \includegraphics[width=4cm,angle=0]{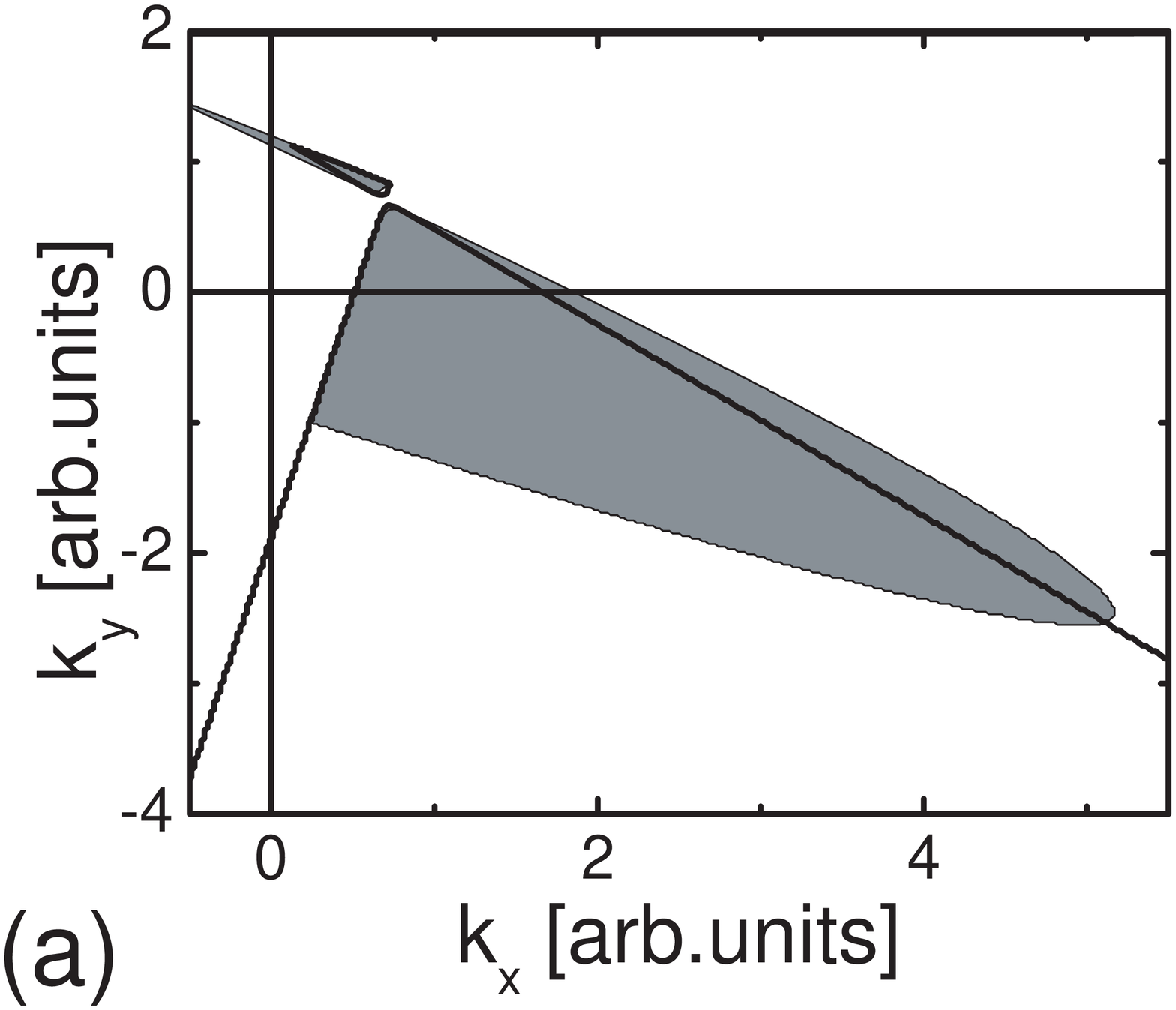} \hfill
\includegraphics[width=4cm,angle=0]{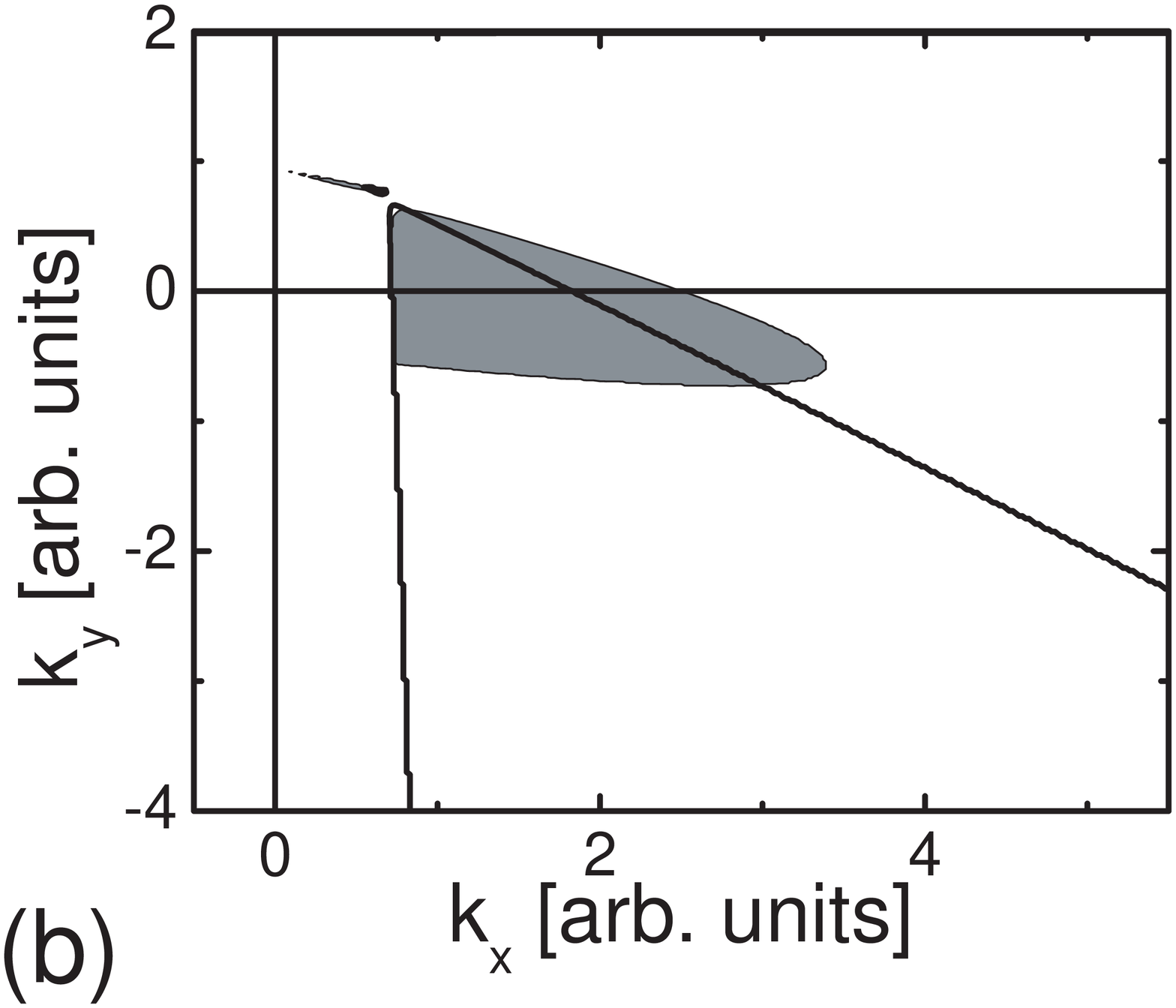}}
    \caption{The domains of stability in the $(k_x,k_y)$ parameter
    plane obtained from the analysis of the model for
    proportional feedback control with a restricted bandwidth at
    (a) rel. pump power $w=1.38$ at a mode configuration $\it [1,2]$ and (b) rel. pump power $w=2.07$
    at a mode configuration $\it [1,4]$. The cutoff
    frequency of the low-pass filter was chosen similar to the
    experimental one to be 300kHz. The lines
    are marking the position of the boundaries of the domains of
    stability for an infinite bandwidth.}
\label{stabdomexpband}
\end{figure}
where $\omega_c$ denotes the cutoff frequency of the low pass filter. Using this form of the
feedback signal and performing the stability analysis we obtain the results presented in
Fig.~\ref{stabdomexpband}. For the simulation a cutoff frequency of 300kHz was used, which
corresponds to the experimental situation. It can be seen that now all essential features of the
experimentally observed region of stability can now be reproduced. This finding shows that an
increased bandwidth gives the possibility to stabilize the laser even further to higher pump powers
up to values important for technical applications.

In conclusion we presented experimental evidence that an intracavity frequency-doubled solid state
laser can be stabilized by a proportional feedback of the two sum intensities of the orthogonal
polarized modes to the pump power. By refining an earlier model it was possible to understand the
shape and dependence on system parameters of the region of stability in the control parameter
plane. The presented result aim in the direction what to do for an application of the method in a
laser with more realistic values of the pump power and, therefore, of the intensity of the light
output. To extend the applicability of the method to a hundred times higher pump powers the
bandwidth has to be increased by the factor ten. This can be easily achieved. We are convinced that
these findings are of great importance for technical application of such a kind of laser.

The authors would like to thank A.~Schenck zu Schweinsberg and U.~Dressler, H.~Kantz, as well as
J.~Parisi for fruitful discussions. K.~Schlenga is acknowledged for his careful reading of the
manuscript and valuable comments. This work was financially supported by the
Max-Planck-Gesellschaft and the Bundesministerium f\"{u}r Bildung, Wissenschaft, Forschung und
Technologie (BMBF) under the Contract No. 13N7036.


\end{document}